**Title:** Measuring Gender and Racial Biases in Large Language Models

Jiafu An[1], Difang Huang[2], Chen Lin[2]*, Mingzhu Tai[2]

[1]Department of Real Estate and Construction, University of Hong Kong

[2]Faculty of Business and Economics, University of Hong Kong

*Corresponding author. Email: chenlin1@hku.hk, jiafuan@hku.hk

**Figures/Tables Record:**

Main Text: 5 figures.

Supplementary Materials: 9 tables and 2 figures.

**Abstract**

In traditional decision-making processes, social biases of human decision makers can lead to unequal economic outcomes for underrepresented social groups, such as women and racial/ethnic minorities.[1-4] Recently, the increasing popularity of Large-language-model (LLM)-based artificial intelligence (AI) suggests a potential transition from human to AI-based decision making. How would this impact the distributional outcomes across social groups? Here we investigate the gender and racial biases of OpenAI's GPT, a widely used LLM, in a high-stakes decision-making setting, specifically assessing entry-level job candidates from diverse social groups. Instructing GPT to score approximately 361,000 resumes with randomized social identities, we find that the LLM awards higher assessment scores for female candidates with similar work experience, education, and skills, while lower scores for black male candidates with comparable qualifications. These biases may result in a 1-2 percentage-point difference in hiring probabilities for otherwise similar candidates at a certain threshold and are consistent across various job positions and subsamples. Meanwhile, we also find stronger "pro-female" and weaker "anti-black-male" patterns in democratic states. Our results demonstrate that this LLM-based AI system has the potential to mitigate the gender bias, but it may not necessarily cure the racial bias. Further research is needed to comprehend the root causes of these outcomes and develop strategies to minimize the remaining biases in AI systems. As AI-based decision-making tools are increasingly employed across diverse domains, our findings underscore the necessity of understanding and addressing the potential unequal outcomes to ensure equitable outcomes across social groups.

**Introduction**

Despite all the efforts by policy makers to promote equal opportunities in the labor market, underrepresented social groups, such as women and racial/ethnic minorities, still face significant gaps in terms of employment rates, income, recognition of attribution, etc[1,2]. Research has shown that social biases of human decision makers in the recruiting process can be a key driver of such inequality in economic outcomes[3-6]. Recently, with the rapid development and increasing popularity of LLM-based generative AI, it is widely expected that a transition from human to AI-based decision making can happen shortly (for example, [link1](link1) show survey evidence that over half of firms are investing in AI-based recruiting). This brings up a new question: if the generative AI is used predominantly to replace human HRs for recruiting, how would this impact the recruiting decisions across different social groups, and how would it change the gender/racial gaps in the labor market?

To date, there is no clear empirical answer to this question. On the one hand, scientists show that large language models such as the GPT make economic decisions with great rationality[7], and generative agents based on fine-tuning techniques can potentially mitigate the biases of large language models[8-10]. Additionally, developers of LLMs have made enormous efforts to restrict the model from providing inappropriate opinions or judgements[11,12]. In addition, anecdotal evidence suggests that popular LLMs such as the GPT are likely pro-democratic and left-libertarian orientated, which suggests that they likely value social equality and diversity (for example, see media reports at [link2](link2) and [link3](link3)).

On the other hand, as the models are trained on human-generated information, generative agents can exhibit similar behavioral biases as humans[13,14], and studies and anecdotes have found significant social biases by earlier-released models[15-18]. Moreover, some debiasing methods may cover biases rather than actually removing them[19]. Thus far, however, such biases are mainly detected in verbal conversations with LLMs, but it is not yet clear whether and how the social stance of these models looks when it is used directly for high-stakes decision making.

In this paper, we quantitatively examine whether and how gender and racial identity affect the LLM's assessment of job candidates. To control for the confounding influences of job candidates' other characteristics that are correlated with their social identity, we utilize an experimental research design that generates a large sample of resumes of entry-level job applicants with randomly drawn work experience, education, and skill sets from real-world distributions of those characteristics. Each resume is assigned a gender and racially distinctive name that reveals the job applicant's social identity. We then instruct OpenAI's GPT, one of the most popular generative AI methods, to assess each randomized resume using a score ranging from 0-100.

Using this randomized experimental research design, we find that, in aggregate, GPT seems to be "pro-social" when screening resumes: it yields significantly higher assessment scores on average for minority (female OR black) job candidates than for white male candidates with otherwise similar characteristics. However, the behavior of GPT varies across different dimensions of social identity: the higher scores for minorities are driven by those for female candidates (both black and white females), but the scores for black males are in fact significantly lower than those for white male candidates. The effects are robust across different job positions and applications from different states. They are also robust when we repeat the analyses many times using randomly drawn subsamples. Meanwhile, we find the "pro-female" behavior to be stronger but the "anti-black-male" pattern to be weaker in democratic states.

Our estimations show that the score differences are not only statistically but also economically significant. Assume that job candidates receiving a score of 80 (out of 100) or above can be hired. According to the GPT's assessments of our sample, this cutoff suggests an approximately 35% probability of being hired for the overall sample. However, our analysis suggests that black (white) female candidates face a 1.7 (1.4) percentage-point greater probability of being hired than otherwise similar white males, while black male candidates face a 1.4 percentage-point lower probability of being hired.

**Generating Resume Scores using LLM**

In our experiment, we developed an automated pipeline using OpenAI's GPT to assess a large sample of randomized resumes for 20 representative entry-level job positions and instruct GPT to generate an assessment score for each of those resumes. The pipeline first parses each job position and the corresponding resumes and then constructs a prompt for the GPT by concatenating a brief, position-specific instruction with the resumes. The prompt is subsequently input into GPT-3.5, the most popularly used and cost-efficient generative AI thus far, generating scores ranging from 0 to 100 for each resume in a single pass. Further details and validation of the pipeline can be found in the Supplementary Information.

To assess whether and how social identity influences GPTs' assessments of job candidates, we systematically evaluate its feedback using a randomized experimental design. Specifically, we generate a sample of approximately 361,000 fictitious resumes with randomized, real-world applicant characteristics, including work experience, educational background, and skill sets (see Methods and Supplementary Information). Each resume is randomly assigned a gender and racially distinctive name to reveal the applicant's social identity. This method follows the most recent labor economics literature[20-22]. If GPT is socially unbiased, the average score for resumes with different social identities should be similar in large samples, given that all the other characteristics are randomized. Instead, if there are differences in average assessments between different social groups, it suggests that the AI might have potential social biases. Comparing the scores of our sample resumes across social groups in Figure S2 in the Supplementary Information (also see detailed results in Table S5), we provide preliminary evidence that GPT is not neutral to social identities: it gives higher scores to female candidates but lower scores to black male candidates.

We then conduct formal regression analyses with an extensive list of controls and fixed effects based on the specification in Equation (1) in the Methods section, although we note that the other characteristics should not confound the results as they are randomized in the sample resumes. Specifically, we control for resume characteristics such as the number of skills, work and educational experiences, the starting date of work and education, the number of high-level work experiences, the average duration of each work experience, the mean word count of work experience descriptions, the total length of education, and the highest educational degree. We also control for position fixed effects, state fixed effects, and fixed effects for the latest work title in each resume. The estimated regression coefficients are displayed in Figure 1 (see detailed results in Table S6 in the Supplementary Information). In the first row, we show that the average score of minority (female OR black) candidates is approximately 0.099 points higher (p<0.05) than that of white male candidates (the benchmark "majority" group). More importantly, in the second and third rows, we find that the average score of female candidates is approximately 0.452 points higher (p<0.001) than that of male candidates, while the average score of black candidates is approximately 0.074 points lower (p<0.1) than that of white candidates. In the last three rows, we further decompose the social groups and demonstrate that, relative to white males, black females score 0.379 points higher, white females score 0.223 points higher, and black males score 0.303 points lower. All these differences are strongly significant, with *p* values less than 0.001. These findings suggest that GPT may assign higher scores to both black and white female candidates but lower scores to black male candidates, conditional on otherwise similar characteristics.

**Robustness Checks**

Figure 2 shows that the differences in scores between social groups remain robust across different types of job positions. First, we compare job positions with higher (above sample median) versus lower (below sample median) shares of female workers according to the national labor statistics from the U.S. Bureau of Labor Statistics (see Table S2 in the Supplementary Information for detailed statistics for each position). As shown in Figure 2a, the scores for female candidates are significantly greater than those for male candidates in both subsamples. Similarly, we compare job positions with higher versus lower shares of black workers in Figure 2b and find that the scores for black male candidates are also significantly lower in both subsamples. The detailed regression results are reported in Table S7 in the Supplementary Information.

In an additional robustness check, we replicate our baseline regression analyses that compare the score differences across different social groups by randomly drawing a 20% subsample each time.

We repeat this random sampling 500 times and summarize the distribution of the estimated regression coefficients in Figure 3. Of the 500 random draws, the mean and median coefficient for black female candidates (which represents the difference in scores compared to white male candidates) are approximately 0.38, and this coefficient is greater than 0.20, with a greater than 99% probability (Figure 3a). A robustly positive distribution of coefficients for white female candidates is shown in Figure 3b, while the coefficient for black male candidates is robustly negative according to the distribution presented in Figure 3c. This robustness check rules out the possibility that our findings are driven by a small number of observations in the sample.

In our baseline prompt to the GPT, we provide a very brief and simple instruction that only specifies the title of the job position. This simple instruction is supposed to simulate an untrained recruiter without any deep knowledge about the position. We also conduct two robustness checks that use representative, detailed job descriptions in the prompt. Specifically, for each position, we use a representative, real-world job posting downloaded from Indeed, a mainstream job search website, and add this job posting to the prompt to the GPT. We consider a version of the prompt with the equal opportunity claim and a version without this claim. As shown in Table S8 in the Supplementary Information, the results remain robust under both versions. More details are provided in the Supplementary Information.

**Heterogeneous Effects by Political Orientation**

The social stance of generative AI could be driven by what it perceives to be the political orientation of the decision-making setting. We examine this conjecture and provide suggestive evidence in Figure 4. As our sample resumes are generated to reflect fictitious job candidates from the five most populous states in the entry-level job market—California, Florida, Georgia, New

York, and Texas—we divide these sample states into two groups based on their political orientation derived from the 2020 presidential election results, such that California and New York are considered Democratic states, while the other three are Republican states. We then compared the score differences across different social groups in each of these two state groups. The results are presented in Figure 4 (refer to Table S9 in the Supplementary Information for detailed results). The first row indicates that the LLM-assessed score of a minority applicant is only significantly higher for candidates from Democratic states. In the second row, we discover that although the scores for female candidates are significantly higher in both state groups, the magnitude is greater in Democratic states. A similar pattern is found when we look at black and white females separately. Moreover, the bias against black male candidates seems to be greater in Republican states. In sum, this analysis suggests that the LLM demonstrates a greater bias toward females but a smaller bias toward black male candidates from Democratic states.

One caveat is that this heterogeneous effect of political orientation should only be considered suggestive evidence, as our sample includes only five states, and we design the prompts without any politics-related information. To better understand the role of political orientation in affecting an LLM's social stance, a more comprehensive research design needs to be adopted.

**Real Consequences on the Probability of Hiring**

We further explore the real consequences of these documented differences across social groups, with a focus on examining the extent to which a candidate's social identity influences his/her probability of being hired. Specifically, we assume that a job candidate can be hired if he/she receives a score from the GPT that is equal to or above a certain threshold. Based on this simple criterion, we can estimate the probability of hiring for each social group under different score thresholds. We consider thresholds of 60, 65, 70, 75, 80, and 85, given that the most frequent scores assigned by GPT are integer multiples of five (see Figure S1 in the Supplementary Information for the distribution of the scores). Under each threshold, we analyze a linear probability model using a similar specification as in equation (1). The dependent variable is a dummy indicator of whether a candidate can be hired under that corresponding threshold, and the key explanatory variable is the corresponding social identity.

Figure 5 shows that the score differences across candidates from different social groups can lead to nonnegligible differences in their probability of being hired. For example, with an 80-point cutoff, which suggests that the average probability of being hired is approximately 35%, black female applicants could expect a 1.7 percentage-point greater probability of being hired than white male applicants, whereas white female applicants have a 1.4 percentage-point greater probability. Intriguingly, black male applicants encounter a 1.4 percentage-point lower probability of being hired than their white male peers who have otherwise similar characteristics. The findings underscore the importance of understanding and addressing the disparities that arise from the use of LLM assessment scores in hiring decisions.

**Discussion**

This paper assesses a key question of debate over the development and prevailing application of LLM: can we alleviate biases against underrepresented social groups by using AI to replace humans in high-stakes decision making? We consider a setting in which social biases of human decision makers have long been documented by social scientists: HR screening of job candidates and hiring decisions. Employing a randomized experimental research design that generates a large sample of resumes with randomly drawn social identity but real-world characteristics in other dimensions (e.g., work experiences, education, skill sets, etc.), we instruct OpenAI's GPT, one of the most popularly used generative AI, to assess each of those randomized resumes and score it. We find that the GPT is robustly biased toward both black and white female candidates but biased toward black male candidates. At a certain threshold, its biases can lead to a 1-2 percentage-point difference in the probability that an otherwise similar candidate can be hired.

This study contributes to the growing literature that examines the potential social biases of LLMs[8,9,15-18]. While most of these existing studies focus on biased or "toxic" expressions in the narratives of AI or on methodologies that can fine tune biases, we consider a close-to-real-world setting from which we can quantitatively infer the social stance of AI in a high-stakes decision-making process.

Our inferences also have important implications for practitioners and policy makers. For practitioners, the application of AI in the recruiting process (and in many other high-stakes processes that traditionally rely on human decision makers) has been increasingly considered a more efficient solution, and the recent development of generative AI, such as GPT, has accelerated this revolution. According to a Gartner survey from June 2023, only 15% of the HR leaders surveyed said they had no plans to add generative AI to their HR processes (Source: [link4](link4)). Moreover, it is not yet well understood whether and how social identities may influence AI decision makers. In fact, there have been concerns that AI may even increase the risk of discrimination in recruiting (for example, see the report of [link5](link5)). On the other hand, there is also anecdotal evidence that some models can sometimes be "too woke" regarding DEI (Diversity, Equality, and Inclusion) issues (for example, see the discussion by Business Insider: [link6](link6)). For policy makers, as the decision-making rule of the AI is a black box, they cannot directly judge from the algorithm whether and how the AI takes the candidates' social identities into its decision making. Using a revealed-preference method, our study provides clear and quantifiable evidence to both parties regarding the social biases of generative AI from multiple dimensions.

This study has several limitations. First, as we focus on the screening of job applications for entry-level positions, for which the job requirements and candidates' backgrounds are much simpler and more homogeneous than those of higher-level positions, our results may have limited external validity and may not necessarily extend to recruiting decisions for high-skill jobs or management positions. Relatedly, to make a clean inference regarding the role of social identities in affecting AI decision making, our sample resumes are designed to follow a standardized format that only includes information about a candidate's work experiences, education, and skill sets. Although all these characteristics are drawn from representative, real-world distributions, the sample results are likely simpler and more homogeneous than the actual ones. Second, our baseline prompts to the AI are designed to include only brief information about the job, and we do not conduct any fine-tuning. Thus, the assessments by AI under our research design can be less sophisticated than those by professional and well-trained human recruiters. Third, our experiment is only conducted at one snapshot of time. As AI continues to evolve and learn, the way it considers social identities in decision making can also evolve.

A potential extension to future work is to better understand how dynamic changes in training sets over time may change the LLM's "social" stance. For example, we have found suggestive evidence that GPT is more "pro-social" when considering candidates from democratic states. As the extent of left versus right leaning in our society is dynamically changing, the information fed to the model and thus the model's social stance may also vary over time. If the relationship between social norms and the social stance of AI can be dynamically mapped, this would shed more light on the social consequences of applying LLM in real-world, high-stakes decision making.

**Methods**

**Experimental Design**

A key challenge in identifying social biases in the recruiting process (and many other real-world activities) is that candidates' social identities can be correlated with other features essential to recruiting decisions[23]. For instance, even when a certain social group appears to have a lower probability of being employed on average, it is difficult to disentangle whether this is due to recruiters' bias against the group identity or because the candidates' skills are less well matched with the specific job.

To address this challenge, we employ an experimental research design that generates a large sample of fictitious resumes with randomized yet real-world applicant characteristics, including educational background, work experience, and skill sets. Each resume is assigned a gender and racially distinctive name to reveal the job applicant's social identity. This method has been widely adopted in recent economic research to detect discrimination in the labor market[20-22]. The underlying concept is straightforward. By considering two large samples of such resumes with different social identities and given that job applicants' characteristics are randomized, the law of large numbers implies that the average quality (and thus the probability of being employed) of the two groups of applicants should be similar if the recruiters who screen the resumes do not exhibit social biases. However, if we observe differences in the recruiters' average assessments (and recruiting decisions) for these two groups of applicants, this outcome would indicate the presence of social biases among the decision-makers. We provide the detailed experimental procedures as follows.

**Resume Creation**

To construct a sample of randomized resumes, we first need a set of representative job positions and real-world job applicants targeting those positions. We begin with a sample of randomly collected entry-level job postings from the mainstream job search websites Indeed and Snagajob. We apply a number of filtering criteria to ensure that those postings were representative (Table S1 presents the distribution of the filtered job postings by industry, state, and position). From this sample of job postings, we select 20 representative job positions with high/medium/low shares of female or black workers based on the corresponding position's national labor statistics reported by the U.S. Bureau of Labor Statistics (BLS). The gender and racial statistics for those 20 selected job positions are reported in Table S2. More details are discussed in Supplementary Note 1.

For each position, we randomly download 1,250 real-world resumes from the job search websites. This gives us a total of 25,000 resumes from which we extract key characteristics, including 1) work experience (e.g., job title, employer name, textual description of job responsibilities, and length of employment for each experience), 2) education experience (e.g., degree title, school name, and length of education for each experience), 3) skills (key words), and 4) the state of residence. We summarize a number of key features of these characteristics in Table S3.

Based on the distributions of those real-world characteristics, we generate a large sample of randomized fictitious resumes. For computational efficiency, we focus on the five states with the highest number of real-world job applicants: California, Florida, Georgia, New York, and Texas. For each position-state pair, we generate a sample of fictitious resumes 40 times larger than the number of real-world resumes we downloaded. For each fictitious resume, characteristics such as work experience, education, and skills are each randomly drawn from the corresponding real-world distributions we obtained earlier. We then randomly assign a gender and racial distinctive name to each resume. In summary, we generated a total of approximately 361,000 randomized resumes, with an equal distribution among the four gender and racial combinations (black females, white females, black males, and white males). More details can be found in Supplementary Note 2.

**LLM Scoring**

We employed the API of ChatGPT-3.5 to evaluate a large sample of randomized resumes. The process is based on an automated pipeline that parses the 20 entry-level job positions and feeds the randomized resumes to each corresponding position. For each resume, the GPT is instructed to provide a score on a 0-100 scale. To ensure the stability of the results, we adjusted the temperature of the ChatGPT API, which controls the concentration of the response (0 being the most concentrated and 2 being the most random). The detailed settings and prompts for the generative AI models can be found in Supplementary Note 3.

**Regression model**

To study the relationship between an applicant's social identity (i.e., whether the applicant belongs to a minority or majority group) and the LLM score for the resume, we estimate equation (1) using a regression specification:

$$Y_{ij} = \alpha_j + \gamma_k + \beta \cdot Identity_{ij} + \boldsymbol{\theta} \boldsymbol{X}_{ij} + \varepsilon_{ij}, \qquad (1)$$

where the dependent variable $Y_{it}$ is the LLM score for the resume of applicant $i$, sent to job posing $j$; $Identity_{it}$ is a set of variables indicating the candidate's social identity, which can be an indicator that equals one if the candidate falls into any minority group (female OR black; in this case, the omitted group is white male), two separate indicators for female (relative to male) and black (relative to white), respectively, or a set of indicators for black female, white female, and black male, respectively (also with white male to be the omitted group). $\boldsymbol{X}_{it}$ is a vector of other resume characteristics, including the number of skills, work and educational experiences, the starting date of work and education, the number of high-level work experiences, the average duration of each work experience, the mean word count of work experience descriptions, the total length of education, and the highest educational degree. In all analyses, we exclude outlier observations with these characteristics falling in the top or bottom 1% of the sample distribution. The summary statistics of these characteristics are reported in Table S4. We also control for job position fixed effects, candidate's state fixed effects, and fixed effects for the title of the applicant's last job. Standard errors are adjusted for potential heteroskedasticity by position. In this specification, our null hypothesis states that the LLM has no social bias when screening job candidates. If the coefficient $\beta$ is significantly different from zero, we can reject this hypothesis.

**Data Availability**

The data that support the findings of this study are available from the corresponding author upon reasonable request.

**Code Availability**

The codes that support the findings of this study are available from the corresponding author upon reasonable request.

**Ethics declarations**

Competing interests

The authors declare no competing interests.

**Supplementary Information**

Please see the Supplementary Materials.

**Figures**

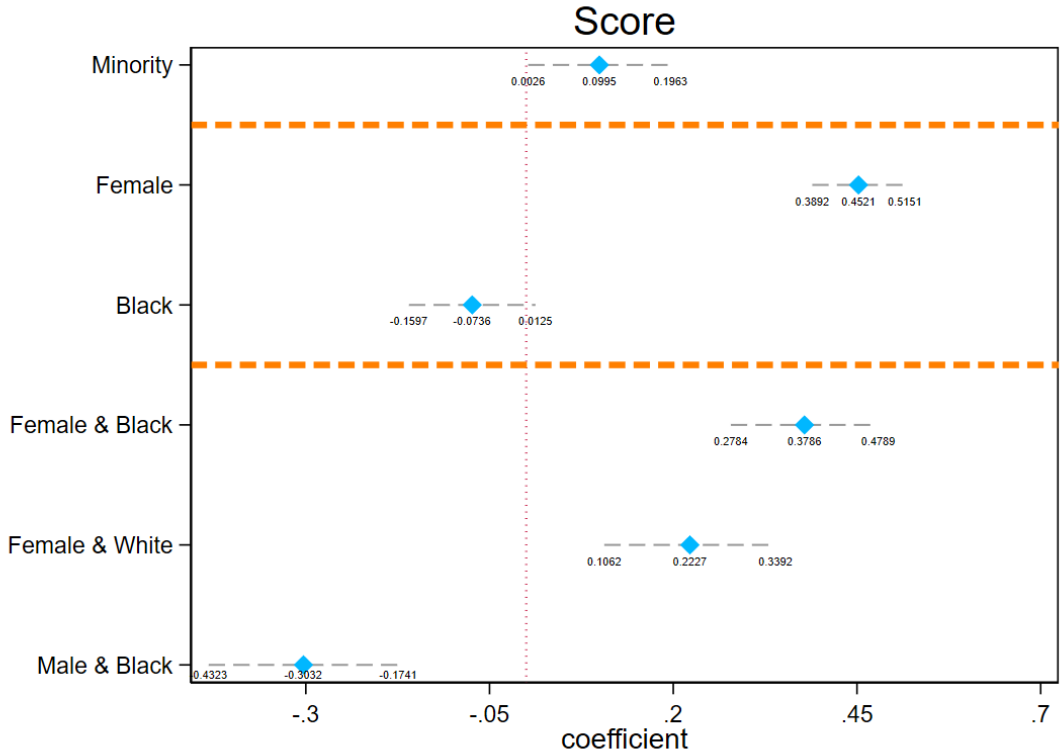

**Fig. 1 | Regression coefficients: score differences across social groups.** This figure presents the regression coefficients that compare the score differences across different social groups, following the specification in equation (1). The dashed orange lines divide the results into three sets depending on different regression specifications. In the first part, we compare the difference in scores between candidates from any minority group (female OR black) and white male candidates (the benchmark "majority" group). In the second part, the comparison is between all females versus all males or between all black versus all white males. In the third part, each minority group is compared with the white male group. The blue diamonds mark the estimated coefficients, and the dashed gray lines show the 95% confidence intervals.

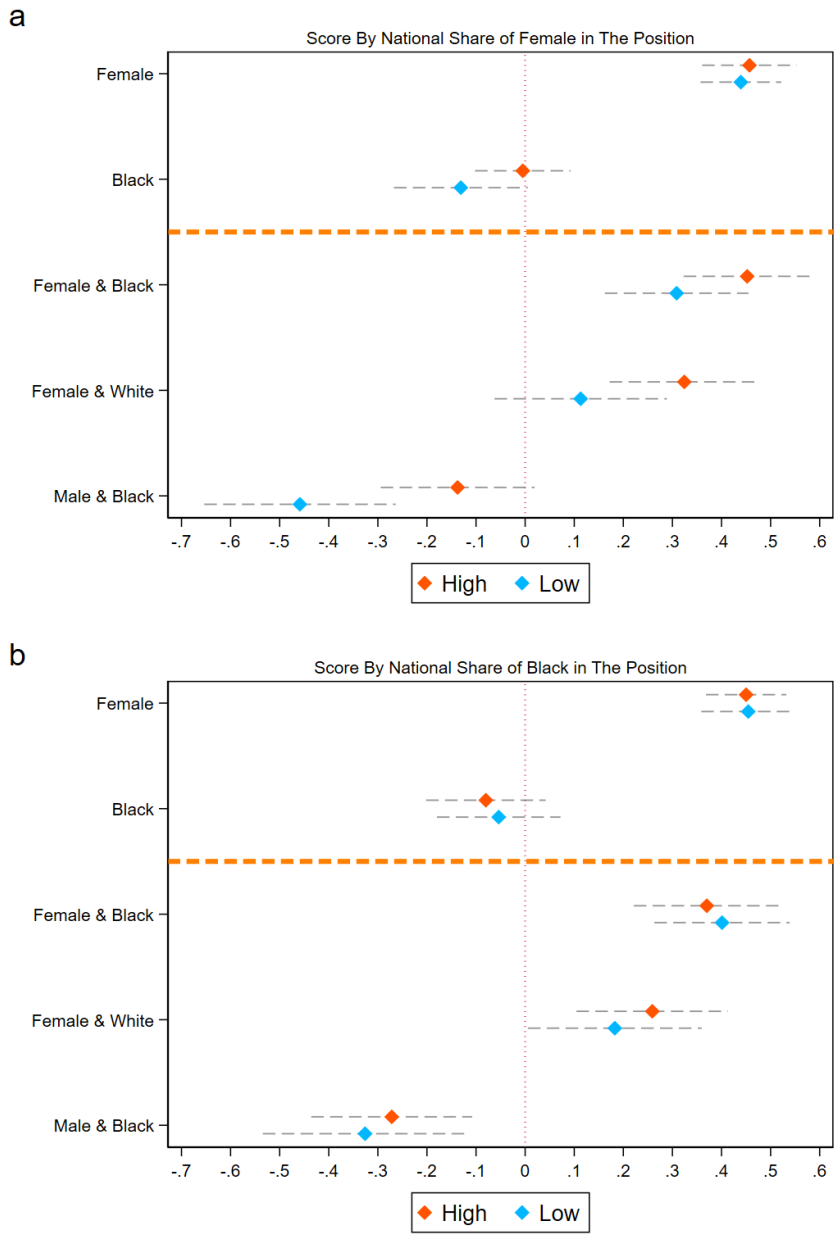

**Fig. 2 | Score differences across social groups by job position type.** This figure compares the regression coefficients, which represent the cross-social group score differences, across subgroups of different types of job positions. In Panel (a), we compare job positions with high (above the sample median, in orange) versus low (below the sample median, in blue) shares of female workers according to the national labor statistics from the BLS. In Panel (b), the comparison is based on the share of black workers. The dashed gray lines show 95% confidence intervals.

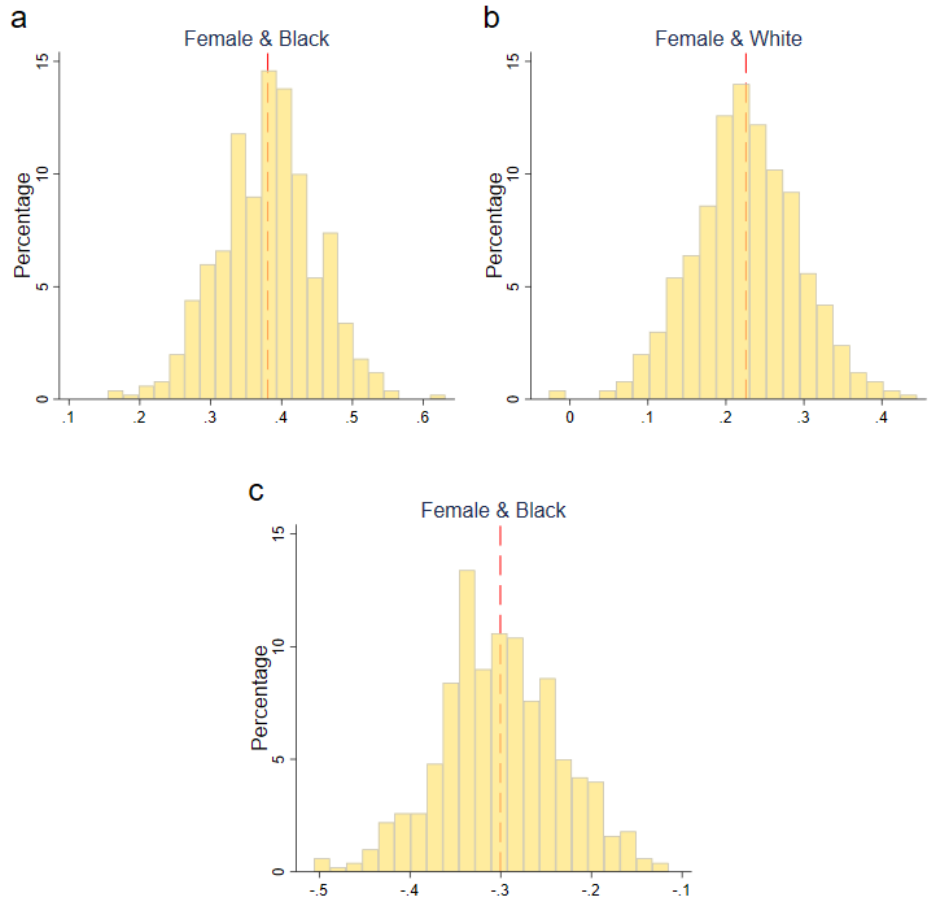

**Fig. 3 | Regression coefficients using random sampling.** This figure plots the frequency distribution of the estimated regression coefficients under random sampling. We replicate the baseline regression by randomly drawing a 20% subsample each time. We repeat this random sampling 500 times and summarize the distribution of the estimated regression coefficients in this figure. The vertical dashed line represents the baseline regression coefficients for the full sample.

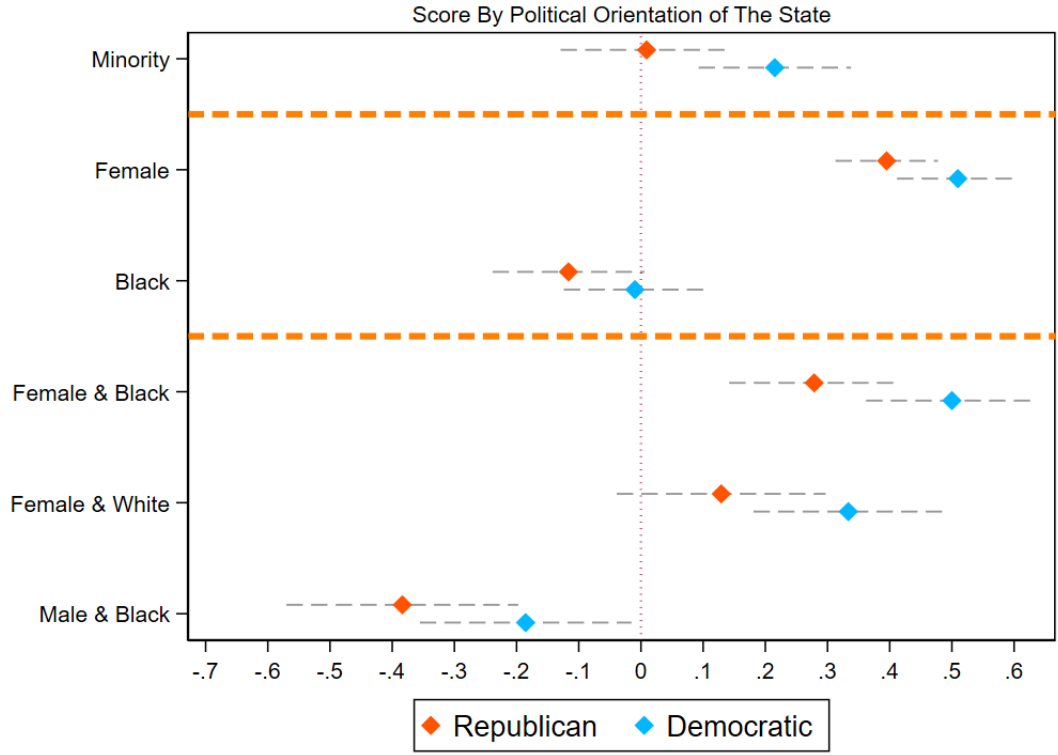

**Fig. 4 | Score differences across social groups: by state political orientation.** This figure compares the regression coefficients, which represent the cross-social group score differences, by states with different political orientations (based on the 2020 presidential election results). Of the five states in our sample (California, Florida, Georgia, New York, and Texas), we consider California and New York to be Democratic states (in orange), while the other three are Republican (in blue). The dashed gray lines show 95% confidence intervals.

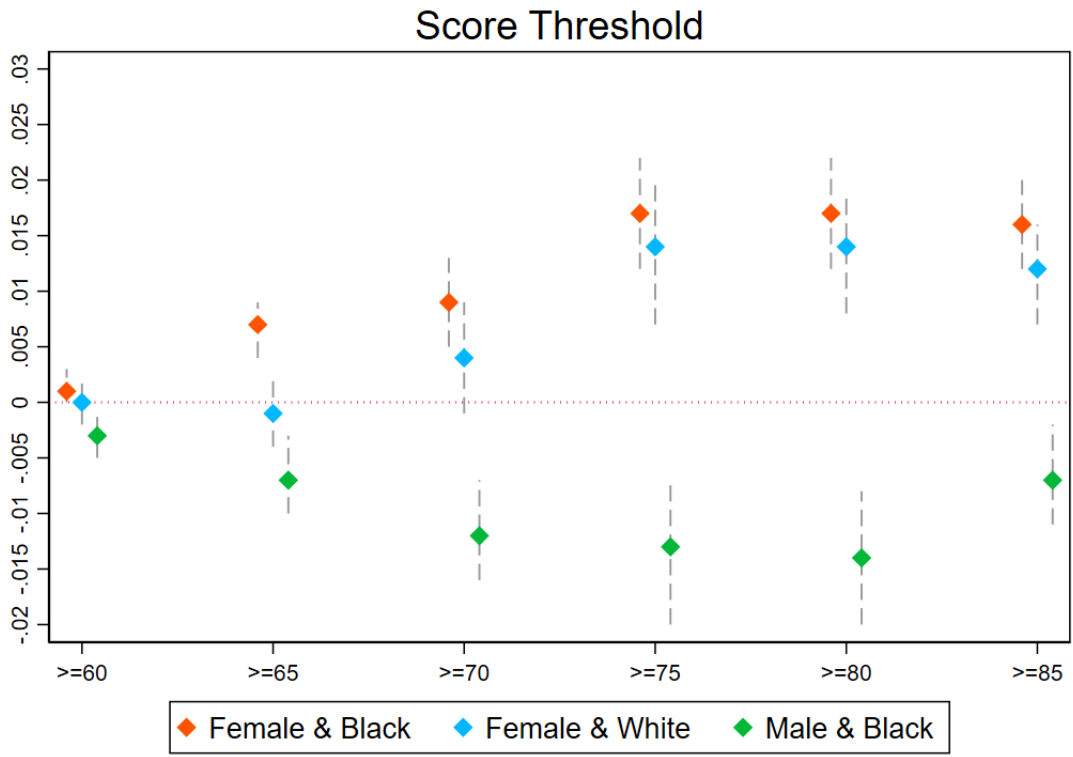

**Fig. 5 | Estimated differences in the probability of being hired.** This figure presents the estimated probability of being hired by a certain minority social group relative to the benchmark group (white male). Under each score threshold, we assume that a candidate with a score equal to or above the threshold could be hired, and we regress the corresponding probability of hiring for each candidate on the social identity using the same regression specification as in equation (1). The regression coefficients are reported in this figure. The dashed gray lines show 95% confidence intervals.